# A Macroscopic Behavioral Violation of No Signaling In Time Inequalities

**Patrizio Tressoldi**

*Dipartimento di Psicologia Genneral*

*Università di  Padova*

*e-mail address: patrizio.tressoldi@unipd.it*

**Markus Maier, Vanessa Büechner**

*Psychology Department, University of Munich, Germany*

*e-mail address: markus.maier@psy.lmu.de; vanessa.buechner@psy.lmu.de*

**Andrei Khrennikov**

*International Center for Mathematical Modeling , in Physics,*

*Engineering, Economics, and Cognitive Science, Linnaeus*

*University, Växjö-Kalmar, Sweden*

*e-mail address: andrei.khrennikov@lnu.se*

## ABSTRACT

In this paper we applied the no-signaling in time (*NSIT)* formalism discussed by Kofler and Brukner to investigate temporal entanglement between binary human behavioral unconscious choices at t1 with binary random outcomes at t2. *NSIT* consists of a set of inequalities and represents mathematical conditions for macro-realism which require only two measurements in time. The analyses of three independent experiments show a strong violation of *NSIT* in two out of three of them, supporting the hypothesis of a quantum-like temporal entanglement between human choices at t1 with binary random outcomes at t2.

**Keywords:** no-signaling in time; temporal entanglement; nonlocal correlation in time; human choices; random events.

# I. INTRODUCTION

The possibility to use mathematical and statistical formalisms adopted in quantum mechanics for the study of biological (e.g., Blankenship & Engel, 2010; Engel, et al., 2007) and cognitive phenomena (e.g., Wang, Solloway, Shiffrin, & Busemeyer, 2014) is not only a theoretical proposal but a rich field of empirical research (see Busemeyer & Wang, 2014; Khrennikov, 2010, for a review).

The application of quantum formalisms to domains other than quantum physics –such as biological or mental processes- is independent to the hypothesis that processing of information by biological systems is based on quantum physical processes within these systems. This approach known as "quantum biological information" is based on the quantum-like paradigm: biological systems of sufficiently high complexity process information in accordance with laws of quantum information theory (Hameroff, Craddock, & Tuszinsky, 2014; Khrennikov, 2010).

However, documenting the usefulness of such mathematical algorithms in modeling decision processes, memory, or consciousness, opens the possibility that the biological substrate constitutes the basis for the emergence of these quantum phenomena. This proposition is controversially discussed and only few researchers share this idea (see e.g., Hameroff & Penrose, 2014). The main argument against the existence of quantum coherence or entanglement in biological systems like the brain refers to decoherence as a strong boundary condition of quantum phenomena (see e.g., Jumper & Scholes, 2014; Tegmark, 2000). Decoherence of quantum states seems to occur with such a high frequency that these effects would be impossible to operate on macroscopically relevant spatial distances or time scales (Tegmark, 2000). This would imply that non-temporal correlations between temporally separated events in the range of several hundreds of milliseconds or even up to seconds would be highly unlikely. In other words, the brain or the parts of it that are involved in actual information processing constitute a macroscopic entity and non-temporal correlations for macroscopic events are quite rare or even impossible (see Tegmark, 2000; but see Hameroff et al., 2014).

Independently of the quantum mind discussion, recently, in psychology, non-temporal correlations between temporally separated events (from a few hundred milliseconds up to several minutes) have been observed (see e.g. Bem et al. 2014; Maier et al., 2014; Mossbridge, Tressoldi, & Utts, 2012). These phenomena usually involved a behavioral or physiological response at time 1 (RP t1) and an activating event happening later at time 2 (AE t2). In these studies a retro-causal influence and therefore temporally non-local correlations of AE t2 on RP t1 were reported.

Since Maier's et al. (2014) studies will be re-analyzed within this article, we will refer to their data in more detail here to illustrate the basic finding. In a series of four out of seven studies a selective key-press at time 1 (left or right) was affected by the random assignment of negative or non-negative picture presentations at time 2. On average the participants were able to avoid negative future events. The random assignment at t2 was performed based on a pseudo random number generator (PRNG) in Study 1, 2, and 3 and with a quantum based random number generator (RNG) in Study 4. In other words the events at t1 and t2 were classically uncorrelated. The findings however indicated that event t1 was affected by event t2 which could only be the case if these macroscopically occurring events were in a state of temporally non-local correlation. Although Maier et al. (2014) reported a significant avoidance effect at t1 being affected by the event on t2, a direct test of temporal non-locality has not been performed. The goal of the data presented here is to fill this gap by providing such a test.

Entanglement in time or temporal non-locality, that is a non-causal correlation between events measured at successive time frames, is one of the many "odd" phenomena studied in quantum physics and mathematical tools have been developed to test the existence of these effects within the empirical data.

Although a commonly accepted mathematical algorithm for a strict test of temporal nonlocality does not exist, some mathematical inequalities that can be applied to temporally distinct physical or mental states have been developed to test the quantum-nature of the underlying physical or cognitive mechanisms. If the inequalities applied to the data are found to be violated, they would indicate the involvement of superposed states.

*Contextual LG inequality and no signaling in time (NSIT) inequality*

The theoretical foundations were originally discussed by Leggett and Garg (1985) as a temporal variant of John Bell inequalities which mainly address entanglement or nonlocal correlations in space. A violation of the Leggett-Garg-equation would confirm quantum-like superposed states between temporally separated events and is thus a pendant of the Bell inequalities for the time dimension. Whereas non-local temporal effects are intensely investigated in quantum physics (e.g. Aharonov et al., 2014; Olson & Ralph, 2012), there are still only few analyses of this type applied to human cognition. Atmanspacher and Filk (2010, 2012, 2013) were probably the first to test temporal non-locality to bistable perception applying their Necker-Zeno model which requires three different measures. Similarly, Asano et al. (2014), derived an analog of the Leggett and Garg inequality, "contextual LG inequality", and used it as a test of "quantum-likeness" of statistical data collected in a series of experiments on recognition of ambiguous figures. The Leggett-Garg approach has some limitations since this test can only be applied for situations involving three consecutively occurring events. For two event scenarios, as is the case in the Maier et al. (2014) research, the Leggett-Garg equation cannot be used. Fortunately, recently a test of non-local correlations for two consecutive events has been developed (Kofler & Brukner, 2013).

*The no-signaling in time (NSIT) inequality.*

Kofler and Brukner (2013), discuss *NSIT* as a further necessary condition to satisfy the Leggett-Garg inequalities to test macro-realism defined by the postulates that a) macroscopic objects which may have two or more macroscopically different states, at any given time, are in a single specific state, b) it is possible to measure this specific state without changing it, and, c) the properties of this macroscopic object are determined exclusively by the initial conditions.

*NSIT* requires only two measurements in time of two dichotomous observables, A and B, that may assume only two distinct states $\pm 1$. Hence, the basic scenario is: $At1= \pm 1$, $Bt1=\pm 1$ and $At2= \pm 1$, $Bt2=\pm 1$.

In accordance with the principle of *NSIT* the outcome probabilities for one part must not depend on the outcome probabilities of the second part and it is expressed by the following formula:

$P(Bt2=+1) = P(At1=-1, Bt2=+1) + P(At1=+1, Bt2=+1)$ and symmetrically

$P(Bt2= -1) = P(At1=+1, Bt2=-1) + P(At1=-1, Bt2=-1)$
(**1**)

A violation of *NSIT* condition could be a first indicator that the mental state evolution cannot be described classically and may be explained by temporally distinct cognitive states existing in a state of superposition.

It is important to note that the temporal nonlocality interpretation of *NIST* is not straightforward and commonly accepted within the scientific community. The most accepted interpretation of violations of *NIST* is that the data that violate these equalities are based on cognitive processes that most likely behaved quantum like. This includes the possibility that the underlying mechanisms are best described as information states that co-exist in a state of superposition. Such a quantum-like behavior of cognitive states could be considered as being a pre-condition for temporal nonlocality to occur. In the analyses presented here we tested this pre-condition. To our knowledge, this is the first attempt to test this formalism in human behavioral tasks.

## II. METHOD

Here we report the analyses of the three formal experiments in Maier et al.'s work (2014), Study 1, Study 2, and Study 4 carried out with participants in the laboratory and with identical conditions and instructions to the participants. Our selection was based on the fact that only in these studies a retro-causal effect of t2 on t1 was observed. One successful study, Study 3, was eliminated since it was completed by a web-based program and participants could not be monitored during their task execution. Thus, only methodologically rigorously obtained data were included. A more detailed description of these experiments is presented in Maier et al. (2014).

*Participants*

In all experiments participants were recruited among the undergraduate and graduate students of the University of Munich, Stony Brook NY, and Barcelona. The number of participants was 111, 201, and 327 for Study 1, Study 2, and Study 4, respectively.

*Procedure*

Each participant was tested individually in a quiet lab room. After the completion of two preliminary tasks, lasting approximately 20 minutes and being unrelated to the crucial study which were devised in order to increase the cognitive fatigue for inducing a more intuitive approach, participants were informed about their new task. A written instruction was presented on the screen: '*In the following experiment you have to press two keys on the key-board as simultaneously as possible. You will see this instruction on the monitor's screen: Please Press the Keys*'. *While seeing this instruction, please press both keys as simultaneously as possible! Afterwards colored stimuli will be presented which you should simply watch.*'

After the participants read the instructions, the experimenter explained that the participants should put their index fingers on the left and right cursor keys of the keyboard. Both keys were placed on the table in front of the participants exactly at the same horizontal position as the midpoint of the computer screen. The experimenter emphasized that both index fingers should slightly touch the cursor keys throughout the experiment, and once the command appears they should press both keys as simultaneously as possible. Participants were informed that there is no rush, but the response should be spontaneous, and that after the key-press they should simply watch the following presentation of a colored stimulus.

Each trial started with the key-press command presented on the screen. Once the key-press was performed, the command line disappeared and, after a 500 msec delay with a black screen, a masked positive (Study 1 or neutral, Study 2 and 4) or negative picture was presented. The masked picture presentation consisted of three consecutive stimulus presentations.

First, a masking stimulus was presented for 72 msec, followed by the presentation of a negative or positive (neutral) picture for 18 msec, again followed by the same mask for 72 msec. Each negative and positive (neutral) picture was combined with an individual mask. The masking stimulus was constructed by dividing the original picture into small squares that were randomly rearranged. The resulting mask consisted of the same color and lightness properties as the original picture and could therefore effectively mask the content of the picture ensuring a subliminal presentation. According to our theoretical model, subliminal perception is critical to allow a superposition of the information states in time. After the second masking stimulus had disappeared, a 3000 msec inter-trial interval appeared before the key-press command initiated the next trial. A total of 60 trial presentations were used in all studies. The 60 experimental trials were preceded by three practice trials with neutral pictures helping the participants to familiarize themselves with the task. Pictures were taken from the International Affective Picture System (IAPS; Lang, Bradley, & Cuthbert, 2008).

Although participants were told to press both keys simultaneously, due to the design of a typical computer keyboard, one of two keys is always triggered first. Thus, in any given trial, either a left or a right key-press was registered even though participants subjectively performed a simultaneous two-key-response. For Study 1 and 2 a closed deck procedure was applied, that is in half of the trials, triggering a left key resulted in a positive (neutral) masked picture presentation and a right key in a negative one. In the other half, key and valence assignment were exactly reversed. The randomization procedure provided by E-Prime™ was used to randomize the order of trial presentation. The 10 positive and 10 negative pictures were randomly assigned to each trial with the restrictions that each picture could maximally be presented 6 times within a study (i.e. if a participant always 'chooses' a positive picture presentation, 60 (6 x 10) positive (neutral) pictures would be presented). In Study 4 an open deck procedure was used, that is the exact assignment to left and right key press and neutral vs. negative picture presentation was abandoned. Also, in this study a quantum-based randomizer, i.e. a true RNG, was used for randomization. Randomized trial selection was performed at the beginning of each trial. After the completion of the 60 trials participants saw each masked picture presentation again and were asked after each whether they could recognize anything and, if so, what.

None of the participants in each of the experiments reported here could precisely name the content of any picture. Thus, the masking procedure met the criterion of subjective unawareness. (from Maier et al.2014, pp. 130-132).

In Study 2 and 4, material, design, and procedure were the same as in Study 1 with the one difference that the 10 negative pictures from Study 1 were used together with 10 neutral instead of positive pictures. Again, the pictures were taken from the International Affective Picture System (IAPS; Lang et al., 2008).

In Study 4, the only difference with respect to Study 2 was that the randomization was obtained by using a quantum based number generator (QRNG) from www.idquantique.com.

*Formal Mathematical Representation*

There are two random variables $A=A_{t_1}$ and $B=B_{t_2}$. The first one corresponds to the first task where the right and left keys determine the values $A=+1$ and $A=-1$, respectively. The nature of another variable is more complicated. The task at $t_2$ determining B is the subliminal perception of a positive or a negative emotional picture. In psychology this task is considered a "response". Now if we assume that these random variables can be represented in the classical probabilistic framework, i.e., there can be introduced the joint probability distribution for their values $P(A=x, B=y)$, the additivity of probability implies that $P(B=y) = P(A=+1, B=y) + P(A=-1, B=y)$.

Typically in applications this equality is treated in the form of the formula of total probability

P(B=y) = P(A=+1) P(B=y /A=+1) + P(A=-1) P(B=y /A=-1).

This formula is violated in a variety of psychological tasks related to disjunction, conjunction and order effects and various probability fallacies (see, for example, Busemeyer & Bruza, 2012; Khrennikov, 2010; Wang et al., 2014).

The main distinguishing feature of the present study is that we couple the violation of the formula of total for statistical data collected in experiments with humans with (non)signaling problem in quantum physics, i.e., time is fundamentally involved into the experimental scheme.

*Application of NSIT formalism*

The left-hand side of equation (1) $P(B_{t2}=\pm1)$ was estimated with a mean equal to 0.5 and a standard deviation of 0.5 assuming a correct randomization.

The probabilities of the right-hand of equation 1, were empirically drawn cross-tabulating the data obtained in the three experiments (see Appendix).

Following the suggestion of Khrennikov et al. (2014), we estimated the standard error of mean (SE) of $P(B\ t2=\pm1)$ taking in account the number of trials of each experiment. The ratio of the observed *NSIT* with the SE was used as an estimate of the *NSIT* violation.

### III. RESULTS

In Table 1 we report the results of the application of the *NSIT* inequality and the standardized deviation with respect to the $P(B\ t2=\pm1)$ in standard errors.

Table 1: Results of the three experiments.

| Study | N trials | SE | NSIT | Δσ |
|---|---|---|---|---|
| Study 1 | 6660 | 0.006126 | **0.068** | **11.1** |
| Study 2 | 8160 | 0.005535 | **0.261** | **47.23** |
| Study 4 | 19611 | 0.003570 | **0.000** | **0.00** |
| Total | 34431 | 0.002694 | **0.0757** | **29.09** |
| Weighted | | | **0.0279** | **10.37** |

SE = standard error of mean; NSIT = no-signaling in time; Δσ = *NIST*/SE

The Δσ values which represent the violation of *NSIT* inequality in term of the number of *SE* from the expected probability at $t_2$, 0.5 in our case, show a clear and strong *NSIT* violation both in the first two experiments and in the analysis of the total trials weighted for the number of trials. It is unclear to us why the *NIST* analysis did not reveal a violation for Study 4. One reason could be the different approaches to realize the trial randomization. Although pseudo random number generators have been used in Study 1 and 2 and a true random number generator was applied in Study 4, PRNG and trueRNG both equally produce random events especially when the seed number and the algorithm used for the PRNG procedure was unknown to the participants, which is the case for our Study 1 and 2. Raw data for independent analyses are available on http://figshare.com/articles/No_Signaling_in_Time_Raw_Data/1383260

### IV. DISCUSSION

Applying quantum mathematical formalisms to test the quantum-likeness of cognitive and behavioral phenomena is becoming more and more popular within the scientific community. In this study we applied the NSIT formalism to investigate temporal entanglement between binary human behavioral unconscious choices at t1 with binary random outcomes at t2. The results of three independent experiments showed a strong violation of *NSIT* supporting the hypothesis of a quantum-like temporal entanglement between the choices at t1 with binary random outcomes at t2 in Study 1 and 2. However, a null result was observed in Study 4. Overall, it seems that for the majority of the data evidence for temporal entanglement could be found. This is to our knowledge the first time that *NIST* formalism has successfully been applied to psychological data sets. Our results therefore support the idea of exploring quantum phenomena within data obtained psychological studies involving unconscious decision making based on automatic affective processes. *NIST* could thus be a valuable tool to test quantum effects in similar paradigms since most psychological experiments consists of activating events and corresponding responses. The main goal of our analyses was to introduce this powerful set of inequalities to a broader psychologically interested scientific community.

In any event, it is too early to be able to draw firm conclusion about the effect of the differences between the studies on the outcome of the *NIST* analysis. At the moment, a pre-registered replication of Study 4 is being undertaken and will be completed in about one year. An additional analysis of these data with *NIST* will shed some more light on the usefulness and applicability of the *NIST* theorem in psychology.

**ACKNOWLEDGEMENTS**

We thank Marco Genovese of INRIM, for an independent analysis of our results and his suggestion on how to weight the total results.

## APPENDIX

Experiment 1

**Key pressed**

| Picture | LEFT+ | RIGHT- | Total |
|---|---|---|---|
| LEFT+ | 1939 | 1391 | 3330 |
| RIGHT- | 1844 | 1486 | 3330 |

P(Q_2= +1)=P(Q_1= -1, P(Q2=+1)+P(Q_1= +1, P(Q2=+1) *NSIT*

0.500=    0.277    +         0.291              -0.0680

P(Q_2= -1)=P(Q_1= +1, P(Q2=-1)+P(Q_1= -1, P(Q2=-1)

0.500=    0.209    +         0.223              0.0680

**Experiment 2:**

|  | **Key pressed** | | |
|---|---|---|---|
| **Picture** | LEFT+ | RIGHT- | Total |
| LEFT+ | 3099 | 981 | 4080 |
| RIGHT- | 3114 | 966 | 4080 |

P(Q_2= +1)=P(Q_1= -1, P(Q2=+1)+P(Q_1= +1, P(Q2=+1) *NSIT*

0.500=    0.382    +         0.380              -0.2614

P(Q_2= -1)=P(Q_1=+1, P(Q2=-1)+P(Q_1= -1, P(Q2=-1)

0.500=    0.120    +         0.118              0.2614

**Experiment 4:**

|  | **Key pressed** | | |
|---|---|---|---|
| **Picture** | LEFT+ | RIGHT- | Total |
| LEFT+ | 4910 | 5024 | 9934 |
| RIGHT- | 4904 | 4773 | 9677 |

P(Q_2= +1)= P(Q_1= -1, P(Q2=+1)+P(Q_1= +1, P(Q2=+1) *NSIT*

0.500=    0.25     +         0.25               0.000

P(Q_2= -1)=P(Q_1= +1, P(Q2=-1)+P(Q_1= -1, P(Q2=-1)

0.500=    0.256    +         0.243              0.000

**Total**

|  | **Key pressed** | | |
|---|---|---|---|
| **Picture** | LEFT+ | RIGHT- | Total |
| LEFT+ | 9948 | 7396 | 17344 |
| RIGHT- | 9862 | 7225 | 17087 |

P(Q_2= +1)=P(Q_1= -1, P(Q2=+1)+P(Q_1= +1, P(Q2=+1) *NSIT*

0.500=    0.286    +         0.289              -0.0754

P(Q_2= -1)=P(Q_1= +1, P(Q2=-1)+P(Q_1= -1, P(Q2=-1)

0.500=    0.215    +         0.210              0.0754